# Integrated Waveguide Brillouin Laser

Sarat Gundavarapu[1*], Matthew Puckett[2*], Taran Huffman[1*], Ryan Behunin[3], Jianfeng Wu[2], Tiequn Qiu[2], Grant M. Brodnik[1], Cátia Pinho[5], Debapam Bose[1], Peter T. Rakich[4], Jim Nohava[2], Karl D. Nelson[2], Mary Salit[2] and Daniel J. Blumenthal[1†]

[1] Department of Computer and Electrical Engineering, University of California at Santa Barbara, Santa Barbara, CA 93106 USA
[2] Honeywell International (United States)
[3] Department of Physics and Astronomy, Northern Arizona University, Flagstaff, Arizona 86011 USA
[4] Department of Applied Physics, Yale University, New Haven, CT 06520 USA
[5] Instituto de Telecomunicações (IT), University of Aveiro, 3810-193 Aveiro, Portugal

*These authors contributed equally to the work
† Corresponding author (danb@ucsb.edu)
(Date 09/12/2017)

## ABSTRACT

**The demand for high-performance chip-scale lasers has driven rapid growth in integrated photonics. The creation of such low-noise laser sources is critical for emerging on-chip applications, ranging from coherent optical communications, photonic microwave oscillators and signal processors to spectroscopy, remote sensing and optical rotational sensors. While Brillouin lasers are a promising solution to these challenges, new strategies are needed to create robust, compact, low power and low-cost Brillouin laser technologies through wafer-scale integration. To date, chip-scale Brillouin lasers have remained elusive due to the difficulties in realization of these lasers on a commercial integration platform. In this paper, we demonstrate, for the first time, monolithically integrated Brillouin lasers using a wafer-scale process based on an ultra-low loss $Si_3N_4/SiO_2$ waveguide platform. Cascading of stimulated Brillouin lasing to 10 Stokes orders was observed in an integrated bus-coupled resonator with a loaded Q factor exceeding 28 million. We experimentally quantify the laser performance, including threshold, slope efficiency and cascading dynamics, and compare the results with theory. The large mode volume integrated resonator and gain medium supports a TE-only resonance and unique 2.72 GHz free spectral range, essential for high performance integrated Brillouin lasing. The laser is based on a non-acoustic guiding design that supplies a broad Brillouin gain bandwidth relative to the cavity resonance width. This leads to robust lasing in the presence of variations in relative offset between the free spectral range and pump-Stokes frequency difference making this a powerful design for high-yield integration. Characteristics for high performance lasing are demonstrated due to large intra-cavity optical power and low lasing threshold power. Consistent laser performance is reported for multiple chips across multiple wafers. This design lends itself to wafer-scale integration of practical high-yield, highly coherent Brillouin lasers with a clear path towards a complete Brillouin laser on a chip including optical pump, spectral filtering, locking resonators and fiber mode couplers and tuning and locking circuits.**

## INTRODUCTION

Brillouin lasers are an attractive resource for a wide range of applications due to their ability to generate highly coherent optical and microwave oscillations and to filter noise in optical signals generated by other sources[1]. These unique features of Brillouin lasers can significantly impact the performance, quality, cost, and energy efficiency of applications including coherent communications, spectroscopy, microwave signal generation, and sensing[2-7]. However, until recently, the exceptional properties of Brillouin laser technology have been difficult to harness due to limitations imposed by discrete component realization and control complexity.

The challenges of creating high-performance Brillouin lasers based on discrete component realization is exemplified in fiber-optic systems. In fiber-based Brillouin lasers, it was demonstrated that exceptional noise characteristics, corresponding to mHz optical linewidths[8] are possible. Yet the realization of such high performance in fiber Brillouin lasers requires increased complexity and cost to compensate for inherent instabilities of discrete fiber-optic and optical component technologies. To overcome these challenges, commercial fiber Brillouin lasers utilize sophisticated assembly and packaging, as well as ingenious feedback and stabilization schemes, to yield reliable state-of-the-art performance (~100 Hz optical line-widths[9]). Monolithic chip-scale integration of Brillouin lasers promises to eliminate many of these challenges. For instance, when pumped with tapered optical fibers, silica wedge-resonators have been used to produce astoundingly low-noise laser oscillation without such technical noise sources[10,11]. These chip-scale devices[10,12,13] have been used to demonstrate compact optical rotation sensors[14], and high-performance microwave frequency sources[7]. However, as the basis for high impact technologies, we require this level of performance from a monolithically integrated Brillouin laser. In this respect, recent demonstrations of Brillouin lasing in membrane-suspended silicon waveguides[15] and chalcogenide waveguides[16] are noteworthy. However, the dynamics of these lasers is quite distinct from conventional Brillouin lasers; further work will be required before these systems produce the extraordinary degree of optical line narrowing that low noise oscillator and gyroscope applications require. The key ingredients for such linewidth narrowing are (1) ultra-low loss waveguides and (2) cavities that support high intra-cavity optical powers.

To reach ultra-high performance with the potential for far-reaching impact, we have developed a Brillouin laser gain medium and resonator from a new class of foundry-compatible waveguide[17,18]. Crucially, these oxide-clad silicon nitride waveguides ($Si_3N_4/SiO_2$)[19] are compatible with wafer-scale material deposition and etching processes, and have losses as low as 0.045 dB/m[18]. The waveguides can be driven to very high-power levels owing to very low nonlinear losses[20] and large effective mode area, making this platform ideal for highly coherent laser systems on a chip. Moreover, these waveguides meet the stringent requirements that, to date, have hindered the creation of fully-integrated Brillouin lasers. Using a $Si_3N_4/SiO_2$ ring-bus resonator waveguide platform, we demonstrate a significant advance in Brillouin laser wafer-scale integration. We fabricated devices on two 4" wafers with seven resonators per wafer and report measurements for nine representative Brillouin resonators. The very low loss single polarization mode waveguides in this platform enable a large mode volume resonator with a TE-only free spectral range (FSR) of ~2.7 GHz, and a loaded Q close to 30 million and intrinsic Q greater than 60 million. The laser performance is tolerant to realistic design and process variations in the FSR relative to the Stokes resonance. This tolerance is due in part to the selectivity of the high Q resonator relative to the large broadened Brillouin gain bandwidth; the latter is a result of non-guided acoustic waves that mediate the Brillouin interaction.

We report the observation of ten Stokes orders and the measurement of input pump vs. Stokes power curves, optical threshold, output power, and slope efficiency for the first three Stokes orders for two optically pumped integrated Brillouin lasers. While previously reported Brillouin resonator designs have required close to exact matching of the cavity free spectral range (FSR) to the 1$^{st}$ order Brillouin Stokes shift, our design is more flexible since the Brillouin Stokes frequency is four times the FSR. The FSR of this resonator is determined solely by the TE mode due to the highly selective (> 75 dB) single TE polarization operation[21,22], a distinct advantage over high Q resonators that support both TE and TM modes.

This $Si_3N_4$ based laser can be directly integrated with a wide range of previously demonstrated active and passive devices including planar waveguides[17], erbium doped lasers[23], arrayed waveguide grating routers (AWGR)[24], sidewall grating filters[25], optical switches and phase tuners[26], optical modulators[27] and switched delay lines[28]. Therefore, a clear path towards a complete Brillouin laser on a chip, using a wafer-scale integration platform, with optical pump, spectral filtering, locking resonators and fiber mode couplers and tuning and locking circuits is within reach.

## LASER CONCEPT AND THEORY OF OPERATION

We create a laser cavity and gain medium by coupling an integrated $Si_3N_4/SiO_2$ bus waveguide and ring resonator as illustrated in Fig. 1(a). We harness the nonlinear acousto-optic coupling in this ultra-low propagation loss waveguide system to create an integrated Brillouin laser by leveraging the properties of a thin silicon nitride core and high purity $SiO_2$ cladding, as reported in more detail in[18] and illustrated in Fig. 1(b). These waveguides support a dilute optical mode, where the mode overlap with the waveguide core is on the order of a few percent and a majority of the mode propagates in the $SiO_2$ cladding. The laser is based on amplification driven by light-sound (Brillouin) interactions[29] of counter-propagating optical fields (pump and Stokes) coupled through a traveling wave acoustic grating schematically depicted in Fig. 1(a). Unique to this chip-scale laser is the fact that the Brillouin coupling is mediated by an acoustic wave that is not guided. Pump light is fiber coupled into the resonator through an integrated bus waveguide. At low input pump power Stokes light is spontaneously generated and as the input pump power is increased the Brillouin gain exceeds the cavity losses and lasing at cavity resonance occurs. As described later in the paper, highly coherent Stokes light is fiber coupled out of the integrated bus and forms the Brillouin laser.

The $Si_3N_4/SiO_2$ waveguide design used to realize the integrated Brillouin laser comprises a high aspect-ratio $Si_3N_4$ waveguide core buried in a $SiO_2$ cladding as depicted in Fig. 1(b). The core material and waveguide geometry is inherently low loss for the TE mode and high loss (>75 dB) for the TM mode[21,22] creating a single polarization mode resonator. This feature is essential in that a pump injection and Stokes frequency generation are defined by a unique resonator passband and FSR without requiring complex dispersion engineering between TE and TM modes and the resulting difficulties in keeping polarization modes aligned or crosstalk between them minimized. The single polarization mode operation of our waveguides is key to realize a uniquely defined free spectral range and single high cavity Q that is the basis to realize a wafer-scale integrated Brillouin laser. The resonator waveguides support a dilute TE-mode as result of the subwavelength core thickness (40 nm). Like fiber, the Brillouin interactions in this waveguide system dominate the optical nonlinearities, enabling highly coherent laser operation on chip. Moreover, the large optical mode area (27.83 µm$^2$), purity of the cladding, and high aspect ratio of the core enable high power handling capability in these waveguides, a quality critical to highly coherent Brillouin lasers[8].

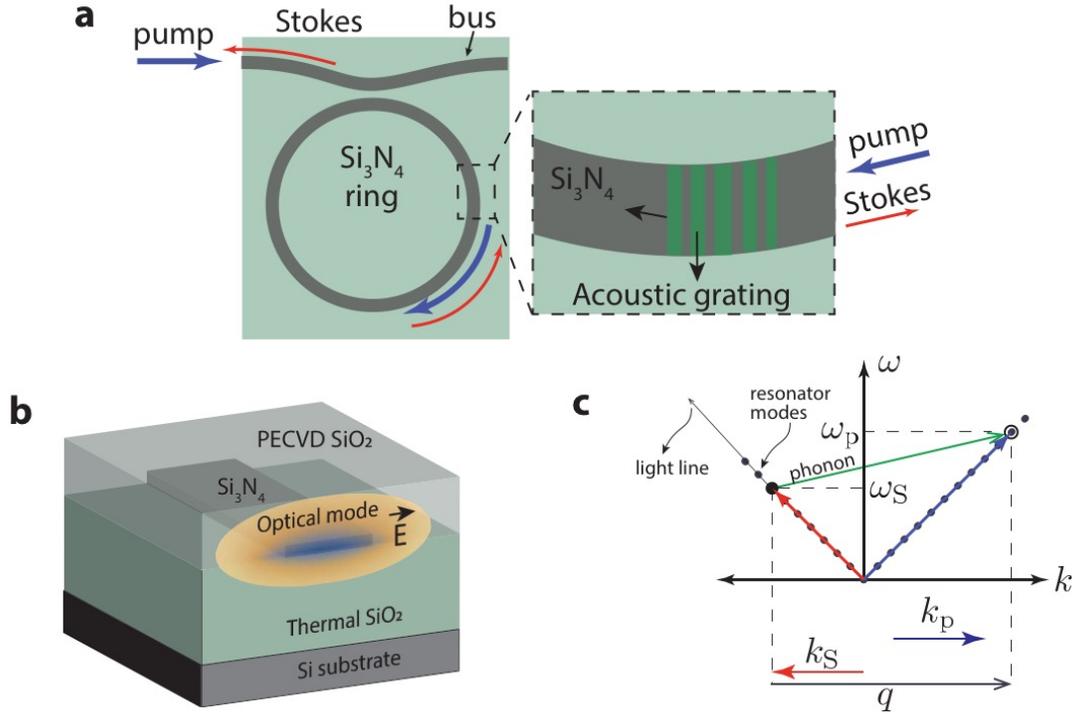

Fig. 1. ***Integrated Brillouin laser based on a confinement-free acoustic and low-loss TE 0ptical waveguide design.*** *(a) Ring-bus resonator configuration and generation of acoustic Brillouin grating and first stokes field from injected pump. (b) Illustration of a $Si_3N_4/SiO_2$ waveguide structure with dilute optical mode that supports low loss TE modes and high TM mode rejection (>75dB). (c) Energy momentum diagram illustrating the ring-bus resonator modes that support the pump frequency and successive backward generated Stokes frequency orders.*

Brillouin scattering is a resonant three-wave interaction involving energy transfer between two optical waves (pump and Stokes) and a travelling acoustic wave[30]. For efficient energy transfer, this interaction requires phase matching between the participating waves, i.e. $\omega_p = \omega_S + \Omega$ and $\mathbf{k}_p = \mathbf{k}_S + \mathbf{q}$, where the ($\omega_p$, $\omega_S$, $\Omega$) and ($\mathbf{k}_p$, $\mathbf{k}_S$, $\mathbf{q}$) are the respective frequency and wavevector of the pump, Stokes, and phonon fields (see Fig. 1(c)). This phase matching condition uniquely relates the Brillouin-active phonon and pump frequencies[30] as $\Omega \approx (2nv/c)\omega_p$, where n is the waveguide effective index and v and c are the respective speed of sound and light. In waveguide structures that accommodate both optical and acoustic guiding such as silica fibers and chalcogenide waveguides[31,32], these phase matching conditions result in a narrow gain spectral bandwidth $\Gamma \sim 2\pi \times 10$ MHz, where this narrow bandwidth is given by the decay rate of the guided phonons. However, our waveguides exhibit a larger bandwidth $\Gamma \sim 2\pi \times 50$ MHz that is desirable for engineerable, robust, highly coherent Brillouin lasers. The absence of acoustic guiding in these waveguides relaxes the phase matching conditions because of the faster phonon decay rate. Additionally, the speed of sound is different in the nitride than in the oxide, further broadening the Brillouin gain. In the presence of a fast phonon decay rate, a weak

Stokes beam can be amplified over a relatively large bandwidth, as it propagates along the low loss optical waveguide.

Lasing is achieved by embedding this Brillouin gain mechanism in the high Q ring-bus resonator with the cavity modes aligned to the pump and Stokes frequencies. As described in the next section, this design has the desired characteristics to create a robust, high performance laser, including large gain bandwidth relative to the resonator linewidth, large resonator mode volume that yields multiple cavity modes per Stokes frequency offset from the pump and supports a large number of cavity photons.

## LASER RESONATOR AND GAIN

A high performance, high yield, wafer-scale integrated Brillouin laser design requires a high loaded quality factor, repeatable bus to ring coupling, single polarization mode operation, high power handling capability, and robust design features such as ease of alignment of the Brillouin gain peak to the resonator. Efficient pump transfer into the cavity is desirable and leads us to incorporate a design with longitudinal modes that can be simultaneously aligned to the pump and Stokes frequencies. Single polarization operation greatly simplifies the design as only one FSR and Q need to be considered for aligning cavity modes to the pump and Stokes frequencies. The combination of a large-volume cavity with single polarization operation ensures that a longitudinal cavity mode is located close to the Brillouin Stokes frequency, greatly relaxing the alignment tolerances and increasing laser chip yield.

The conceptual relation between the integrated resonator and the pump and first order acoustic and optical Stokes waves is illustrated in Fig. 2 showing the resonator reflection spectra and FSR (black), the optically generated acoustic grating and resulting Brillouin gain spectra $G(\nu)$ in a non-acoustically guided structure (green), the input pump signal (blue) and the resulting first Stokes order (red). One of the key advantages of our platform is a TE-only high-Q cavity with multiple FSR resonances per Brillouin shift $\Omega$. The frequency spacing between the resonator's optical modes is given by $c/n_g L$, where $n_g$ is the group index of the optical mode and L is the resonator length.

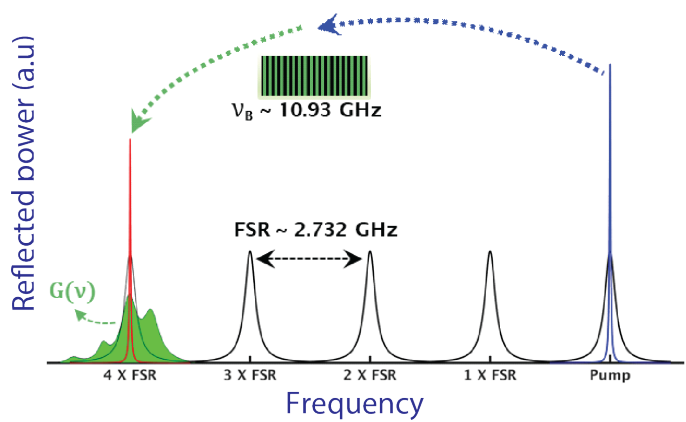

Fig. 2.  *Conceptual illustration of integrated Brillouin lasing.* Shown are resonator reflection spectra and the 2.732 GHz FSR (black), optically generated acoustic grating and resulting broadened Brillouin gain spectra $G(\nu)$ in a non-acoustically guided structure (green), input pump signal (blue) and generated first Stokes order (red).

The group index was measured to be $n_g = 1.478$ using an optical backscatter reflectometry technique[18] with test structures. In order to estimate the Brillouin Stokes shift, we performed a calibrated simulation (described in the methods section) based on full vectorial optical and acoustic finite element simulations[29] that incorporated measured waveguide material acoustic properties. These calibrated simulations guided our resonator design process by determining the Brillouin frequency and the absolute Brillouin gain.

The gain profile is obtained using pump probe measurements on a 5 m waveguide spiral structure. A key characteristic shown in the simulated and measured Brillouin gain curves in Fig. 3(a) is the significant broadening due to the absence of acoustic waveguiding in our laser resonator. This broadening, on the order of 200 MHz, leads to robust lasing operation and fabrication. The calibrated gain simulations were critical in guiding our initial test structure design and gain measurements. These simulations were corroborated by the measured gain profile and elucidated the nature of the acoustic modes participating in Brillouin scattering process in the absence of acoustic guiding.

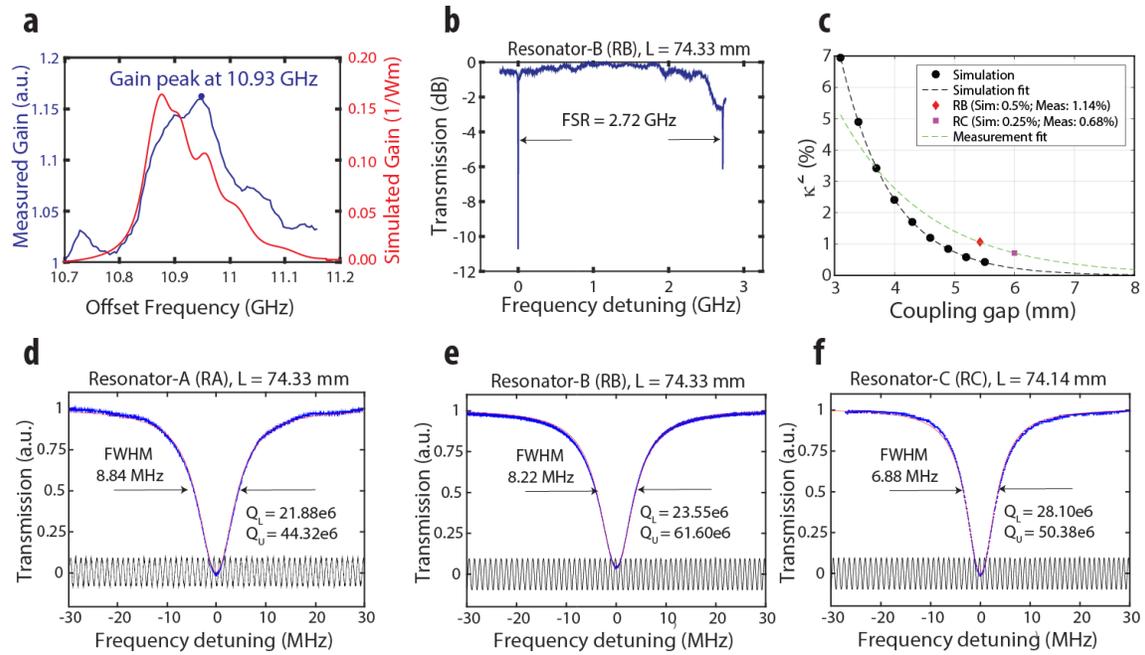

Fig. 3. ***Simulations and measurements of Brillouin gain and laser resonator coupling, Q and FSR.*** *(a) Simulated and measured Brillouin gain spectra from a test non-acoustically guiding TE only waveguide. (b) Representative transmission spectrum showing the free spectral range of 2.72 GHz, demonstrating a unique FSR due to single polarization operation. (c) Simulation and measurement of bus-ring coupling coefficient for two resonators from two different wafers. (d), (e) and (f) RF calibrated MZI measurements of transmission FWHM and calculated quality factors using Lorentzian fit to transmission spectra for resonators RA, RB and RC respectively.*

We designed the resonator FSR to be ¼ the Brillouin frequency shift ($\Omega = 10.93$ GHz), which is possible in a large mode volume resonator, in order to guarantee that the Brillouin-active optical mode is within 1.366 GHz from resonance. We combine this large-mode volume FSR with broad Brillouin gain (due to the absence of acoustic mode waveguiding and different acoustic velocities in the nitride and oxide) in order to realize a fabrication robust design. We fabricated two different resonators with physical lengths of 74.3 mm and 74.1 mm corresponding to desired FSRs of 2.728 GHz = (10.91 GHz)/4 and 2.732 GHz = (10.93 GHz)/4 respectively. The bend radii of both resonators, 11.83 mm and 11.80 mm respectively, were chosen to be larger than the critical bend loss limit for this waveguide design. The FSR of a 74.33 mm long resonator was measured to be 2.72 GHz FSR as shown in Fig. 3(b). This measurement was made using a single sideband tunable swept optical source and differs by only 8 MHz from the design FSR, a small fraction of the ~ 200 MHz Brillouin gain bandwidth.

Once we designed the resonator FSR relative to the Brillouin Stokes frequency shift, the laser performance is strongly defined by the resonator quality factor that is comprised of both loaded ($Q_l$) and the unloaded ($Q_u$) components. The balance between $Q_l$, determined by the waveguide and bus-ring coupling losses, and $Q_u$ determined by the waveguide losses, leads to design tradeoffs including optical threshold, output power and coherence properties as well as power efficiency. On one hand, a large $Q_l$ is desirable because it yields a larger buildup of power within the resonator for a given on-chip pump power, lowers the laser threshold, and improves the laser phase noise and hence the linewidth. However, this performance boost comes at the cost of reduced laser emission and wall plug efficiency among other factors.

In order to realize high Q resonators, we used test structures to calibrate simulated coupling coefficients of fabricated couplers. Simulation of the coupling coefficient as a function of ring-bus gap was performed using PhoeniX Software's Optodesigner and is shown Fig. 3(c) (the black curve). Resonators with gap values of 5.42 μm and 6.00 μm correspond to the simulated coupling coefficients of 0.5% and 0.25% respectively, as indicated by the data points in Fig. 3(c). Measured values of these test structures returned coupling coefficients of 1.14% and 0.68% for gap values of 5.42 μm and 6.00 μm (red data points on the green fit curve).

The quality factor was measured using an RF calibrated Mach-Zehnder interferometer (MZI)[33], as shown in Fig. 3(d-f). For the 0.71% coupled resonator, the measured FWHM was 6.88 MHz corresponding to a loaded Q close to 30 million and an intrinsic Q of over 50 million (using a Lorentzian fit and assuming negligible coupler excess loss). For the 1.14% coupled resonator, the FWHM was measured to be 8.22 MHz, corresponding to a loaded Q of 23.5 million an intrinsic Q of 61.6 million. Using the same Lorentzian fit model, propagation losses of 0.52 dB/m for the former resonator and 0.42 dB/m for the latter resonator were extracted. A complete set of measured Q values for 9 representative resonator chips across two wafers is presented in the following section. The derived waveguide propagation losses correspond closely to measurements of our test linear spirals using optical backscatter reflectometry as discussed in the methods section.

## LASER OPERATION AND CHARACTERIZATION

Operation of our integrated Brillouin laser demonstrating cascaded lasing of ten Stokes orders (S1 – S10) is shown in Fig. 4 and Fig. 5. Using the apparatus illustrated in Fig. 4 we measured the laser dynamics. A fabricated integrated Brillouin laser chip is shown in in Fig. 5(a) with dimension 25mm x 27.5mm. Measurements included transmission and reflection optical spectra using an OSA, RF beat tones between cascaded Stokes orders using an ESA, Stokes power transfer curves at both ports, and optical threshold for three Stokes orders of devices from two different wafers. A 1550nm tunable fiber pump laser was coupled to the resonator mounted on a temperature controlled stage. The pump was locked to a resonator mode using an external phase modulator and Pound-Drever-Hall (PDH)[34] feedback loop as shown in Fig. 4 and described further in the methods section.

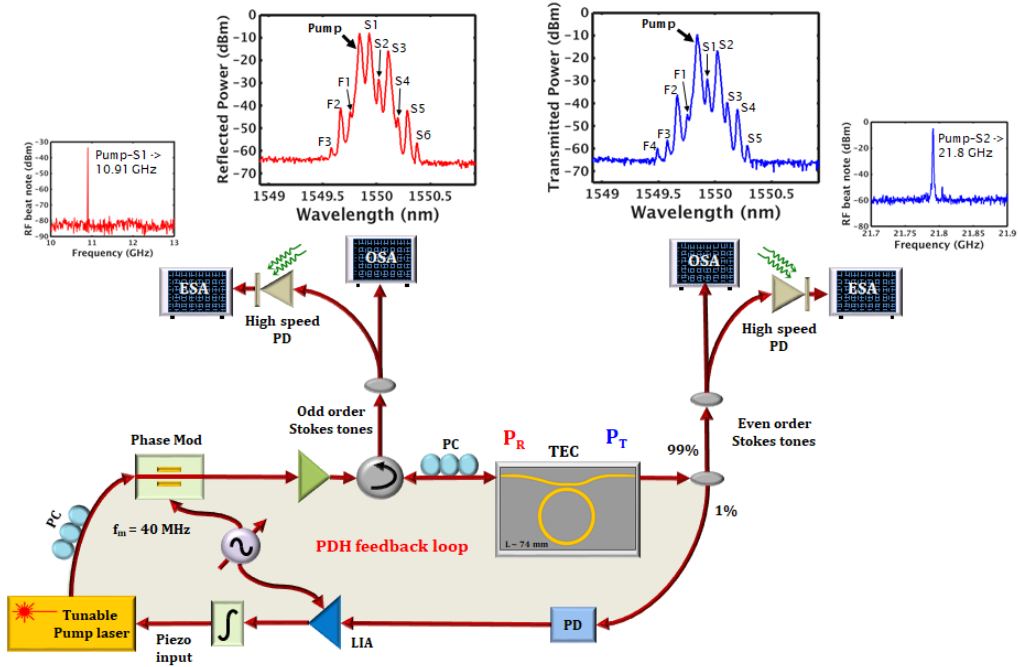

Fig. 4.  **Brillouin laser experimental setup.** *Illustrating integrated Brillouin gain medium and resonator chip on temperature controlled mount, with external pump laser and PDH feedback loop. OSA and ESA measurements at the reflection and transmission ports were used to measure Brillouin even and odd Stokes orders and pump-S1 and pump-S2 RF beat-tones.*

The coherent emission of Stokes light is measured at the reflection port ($P_R$) and transmission port ($P_T$) for the odd and even Stokes orders respectively. Ideally, as Stokes energy buildup occurs, the odd Stokes orders are generated counter-propagating to the pump and are emitted from port $P_R$, while the even Stokes orders co-propagate with the pump and are emitted from port $P_T$. However, reflections at the chip facets result in energy from a portion of odd orders appearing at port $P_T$ and a portion of energy from even orders appearing at port $P_R$. An example of transmission and reflection optical spectra and an RF beat tones are shown in the OSA and ESA traces in Fig. 4.

Brillouin laser emission for the first Stokes order (S1) is observed (see Fig. 5(b)) as a function of the on-chip pump power. As an example, the onset of lasing for Resonator-B (RB) was observed at ~50 mW as the pump is increased. As the pump power is further increased the cascading of higher Stokes orders is observed and the power in S1 becomes clamped. Using an optical spectrum analyzer to measure the output of both bus coupled ports we observe the pump and the even and odd Stokes orders as shown in Fig. 5(c). Cascaded Stokes order lasing dynamics become evident when the pump power is increased well above the lasing threshold which causes the circulating Stokes power to reach sufficient levels to initiate lasing in higher order SBS modes demonstrating highly efficient cascaded lasing up to 10 Stokes orders (S1 – S10). The spectra in Fig. 5(c) is taken at pump input power well above threshold.

In addition to optical fields generated from the Brillouin nonlinearity, four-wave mixing (FWM) produces frequencies that coincide with Stokes orders and generates new frequencies on the blue shifted side of the pump, similar to observations in Chalcogenide waveguides[35]. We indicate the FWM frequencies by $FWM_{X;Y;-Z}$ as shown in Fig. 5(d), where X, Y and Z are the contributing Stokes orders. This contribution of FWM is observed in Fig. 5(d). The pump power is initially set to generate only S1 and FWM is not observed (grey trace). As the pump power is increased to generate S2, degenerate FWM between the pump and S2 is observed as the blue shifted and red shifted tones denoted by $FWM_{P;P;-S2}$ and $FWM_{S2;S2;-P}$ in Fig. 5(d). The blue shifted FWM field cannot be explained as an anti-Stokes process and the red shifted FWM field cannot be efficiently generated by Brillouin scattering until S3 reaches threshold.

Lasing threshold and pump vs. Stokes power were measured for two sample resonators and three stokes orders. The optical pump vs. first order Stokes power is plotted for resonators RC and RB in Fig. 5(e) demonstrating threshold at 12.7 mW and 34.1 mW respectively. The loaded Q values of 28.10 million and 23.55 million for these resonators demonstrate the increased efficiency and decreased threshold with increased loaded Q.

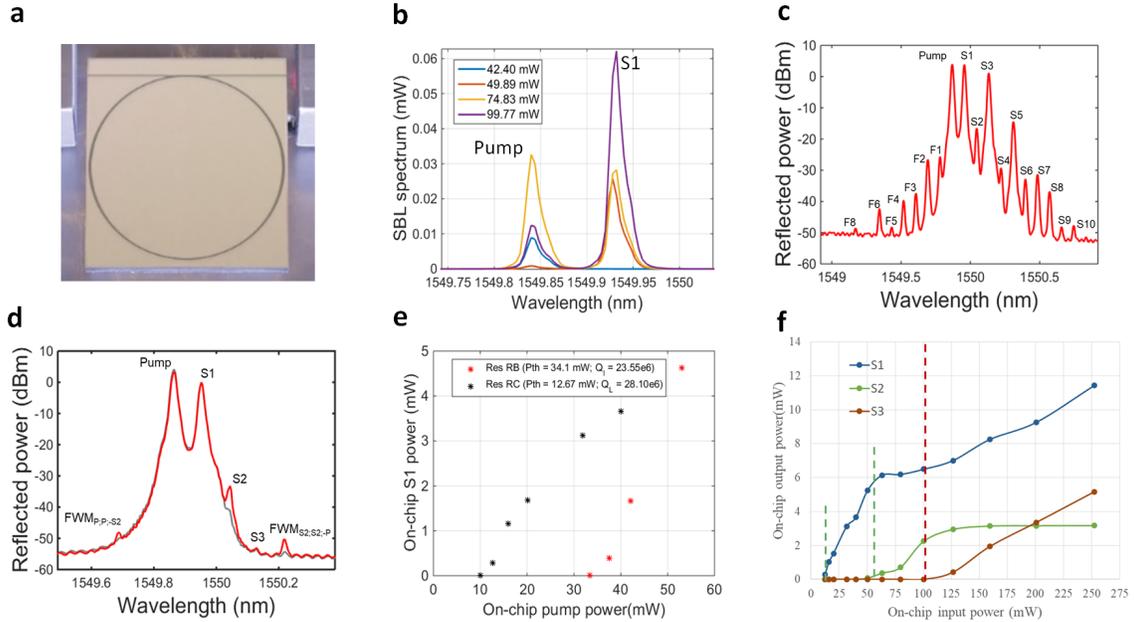

Fig. 5. ***Demonstration and measurement of an integrated Brillouin laser.*** *(a) Contrast enhanced photograph of silicon nitride laser chip. (b) Threshold behavior of first Stokes order for resonator RB. (c) Reflection port optical power spectrum for resonator RA showing cascading of 10 Stokes orders and four wave mixing tones. (d) Demonstration of four wave mixing generation. (e) On-chip pump power vs. on-chip first Stokes order power for resonators RB and RC. (f) On-chip pump vs. on-chip Stokes power for first three orders, illustrating multiple mode laser dynamics including cascaded power transfer.*

By lowering the coupling factor to 0.25% and increasing the loaded Q to 28 million, the threshold power for resonator RC decreased to 12.7 mW with a slope efficiency close to 7%. The decrease in slope efficiency for RC occurs due to various factors including power shedding into higher Stokes orders. The threshold optical power for S2 and S3 for resonator RC were measured to be approximately 50 mW and 100 mW respectively. The on-chip optical powers for three Stokes orders, S1, S2 and S3, is shown in Fig. 5(f) as the input pump power is increased. Power clamping due to cascaded lasing is observed, as shown in Fig. 5(f), for S1 at the onset of S2 lasing and for S2 at the onset of S3 lasing.

This laser technology is compatible with commercial 4" silicon nitride wafer-scale processing[17]. Our lasers were fabricated on thermally grown oxide on silicon 4" wafer substrates with 7 lasers per wafer as shown in Fig. 6. To characterize the repeatability of resonator quality across wafers we measured the loaded Q factor ranging from 21 million to 31 million, for representative die as shown in the table in Fig. 6. Also shown are the optical threshold power for lasers across two wafers.

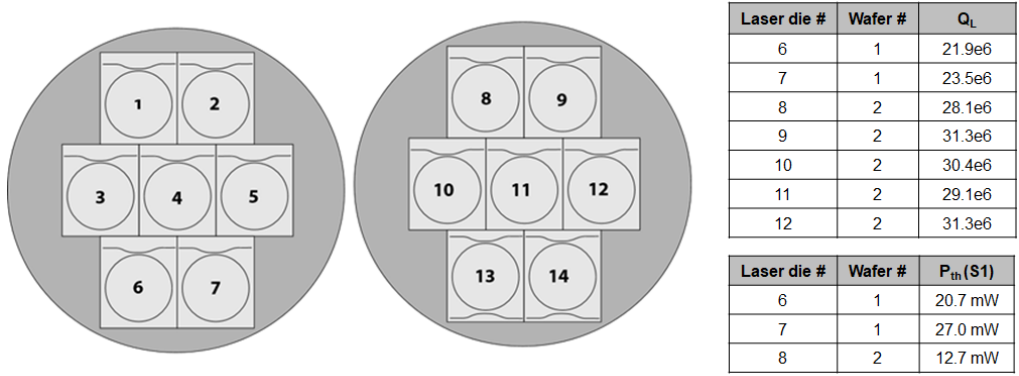

Fig. 6. *Wafer-Scale Measurements of Brillouin Laser Die.* Two wafers were fabricated with a total of 14 die. Laser die number L6, L7 and L8 correspond to passive resonator measurements RA, RB and RC. Loaded Q values are measured using the RF calibrated MZI technique described in the methods section. Threshold values for first Stokes order are shown with lowest value of 12.7mW.

## DISCUSSION

We have demonstrated a wafer-scale integrated Brillouin laser consisting of an SBS gain medium and a bus-coupled resonator. By combining Brillouin lasing physics derived from low-loss optical waveguides and unguided acoustic modes, we achieve world-class performance compatible with a commercial $Si_3N_4$ foundry platform. Owing to these properties, this Brillouin laser is poised to create high technological and scientific impact and open paths to low cost, compact, highly coherent, low noise laser systems on a chip. The ability to create more complex photonic circuits with previously demonstrated active and passive components in this platform promises to impact a wide range of applications including communication, RF microwave sources, sensing and navigation.

Stimulated Brillouin lasing to 10 Stokes orders was observed in a resonator with a loaded Q factor exceeding 28 million. Optically pumped threshold values of 12.7 mW and 34.1 mW for first Stokes orders were measured for loaded Q values of 28.10 million and 23.55 million respectively. The large mode volume integrated resonator and gain medium supports a TE-only resonance and unique 2.72 GHz free spectral range, essential for high performance integrated Brillouin lasing. The laser is based on a non-acoustic guiding design that supplies a broad Brillouin gain bandwidth relative to the cavity resonance width. This leads to robust lasing in the presence of variations in relative offset between the free spectral range and pump-Stokes frequency difference making this a powerful design for high-yield integration. Laser dynamics were measured from lasers from two different wafers, and passive resonator characteristics measured for nine of the 14 laser die, demonstrating wafer-scale compatible integration.

Looking forward, a number of design strategies are available to drastically improve the performance of this laser system. For example, the resonator Q can be engineered by altering the waveguide design and implementing new processes to further lower waveguide loss. The coupling losses can be minimized by increasing the resonator length, and sidewall scattering can be reduced by adjusting the waveguide width and optimizing the post-fabrication annealing process. In addition to design and processing adjustments just described, the laser dynamics can be further engineered by manipulating the frequency response of the resonator as well as the Brillouin gain. Utilizing these two strategies, laser cascading can be controlled so that the power and coherence of individual Stokes orders can be adjusted.

By adopting these strategies, we foresee that resonators with loaded Qs in excess of several hundred million and engineered Brillouin emission will enable unprecedented coherent integrated lasers. These lasers will possess linewidths approaching mHz and phase noise properties previously achievable only with non-integrated technologies.

## ACKNOWLEDGEMENTS

This material is based upon work supported by the Defense Advanced Research Projects Agency (DARPA) and Space and Naval Warfare Systems Center Pacific (SSC Pacific) under Contract No. N66001-16-C-4017. The views and conclusions contained in this document are those of the authors and should not be interpreted as representing official policies of DARPA or the U.S. Government. We would like to thank Robert Lutwak and James Adeleman for useful discussions. We also thank Biljana Stamenic for help in processing samples in the UCSB nanofabrication facility, William Renninger for help with the measurement techniques for Brillouin gain profiles and Joe Sexton, Jim Hunter and Dane Larson at Honeywell for the cladding deposition, pre-cladding preparation and anneal process.

## DISCLAIMER

The authors declare no competing financial interest.

## CONTRIBUTIONS

S. G., R. B, P. T. R. and D. J. B. prepared the manuscript. S. G., M. P. and J. W. contributed equally to performing the system measurements. T. H., D. B. and J. N. contributed to the silicon nitride integrated laser fabrication. M. P., D. B., J. N., K. D. N., M. S and D. J. B. contributed to the laser design. R. B., P. T. R., M. P., T. Q., S. G., and K. D. N. contributed to the calibrated Brillouin simulation and modeling. G. M. B., C. P. and S. G. built the RF calibrated and measure the Q values of laser resonators. III and III contributed to the device characterization and gain measurements. All authors contributed to analyzing simulated and experimental results. D. J. B., K. D. N., P. T. R., and M. S. supervised and led the scientific collaboration.

## Methods

**Fabrication methods:** A 40 nm thick silicon nitride film was deposited using LPCVD on a 4-inch silicon wafer with a 15 μm thermally grown oxide. The wafer was spun with standard DUV anti-reflective (AR) and photoresist layers and then patterned using an ASML PAS 5500/300 DUV stepper. The AR coating was etched with a RIE PlasmaTherm etch tool. The resist was used as an etch mask to realize high aspect ratio waveguide core by anisotropically dry-etching the silicon nitride film in a Panasonic E640 ICP etcher in a $CHF_3/CF_4/O_2$ chemistry. The resulting byproducts from this etch were then ashed with a Panasonic E626I ICP tool in an O2 atmosphere before stripping the resist by sonicating in a hot NMP solution and rinsing in Iso-propanol. The AR coating and other organic impurities were then removed by dipping the wafer in a freshly prepared, standard piranha solution heated at 100 °C. The wafer was inspected by dark-field microscopy for undesired particles near the waveguide core at both post-develop and post-etch stages and additional sonication in hot NMP solution and rinse in Iso-propanol were performed if required. An additional plasma clean using a Gasonics Aura 2000-LL Downstream asher tool helped us get rid of any leftover resist and other organic materials. The device features were inspected with a JEOL 7600F FE-SEM and the RMS sidewall roughness was measured using Dimension 3100 AFM to be lower than 3 nm. A 6 μm thick oxide was deposited using plasma enhanced chemical vapor deposition with TEOS as a silicon precursor followed by a two-step anneal at 1050 °C for 7 hours, and 1150 °C for 2 hours. The wafer was diced in to individual devices (2 spirals per wafer/7 resonators per wafer) to be used for Brillouin gain characterization and lasing respectively.

**Calibrated simulation methods:** Prior to fabricating our Brillouin lasers, we analyzed the optical and mechanical properties of the materials used in our waveguide structure. The refractive index, material density, and Young's modulus of thermally grown oxide, LPCVD nitride, and PECVD oxide films on Silicon wafers were measured using optical ellipsometry, X-ray reflectometry, and nano-indentation techniques respectively. With the parameters evaluated from these studies, we theoretically predicted the position, shape, and amplitude of the SBS gain coefficient of our waveguides. To do this, we constructed fully three-dimensional finite element method (FEM) models of the waveguides to study their acousto-optic behavior. The simulated electromagnetic field as supported by our waveguide structure was used as a dipole source to drive acoustic waves. The power contained in the acoustic wave was calculated as the acoustic frequency was parametrically swept across the analytically predicted phase-matched frequency. The resulting shape of the acoustic wave was found to form at the frequency corresponding to the gain peak at 10.92 GHz. The simulated gain spectrum as shown in Fig 3(a), exhibits a high degree of skewness toward higher frequencies, which is a result of the lack of acoustic guidance, that causes acoustic waves to couple into a portion of the free-space continuum. The predicted degree of broadening is consistent with the value estimated through observation of the angle of divergence of a single acoustic wavelet. The estimated Brillouin gain spectrum and gain coefficient were used to estimate the threshold of first three Stokes orders using the coupled mode equations of the pump and Stokes tones. More details about the simulation of Brillouin gain achieved in a non-acoustic guiding $Si_3N_4$ waveguide can be found in the supplemental section.

**Waveguide loss and coupler measurements:** A 5 m in/out waveguide spiral was used to measure the Brillouin gain spectrum of this waveguide geometry. The spiral consists of 42 turns stitched between four DUV masks. This means there are 168 stitches in total. The spiral has a central s-bend where it reverses spiraling direction. This s-bend has the smallest radius in the design of 11.83 mm. The propagation loss of the spiral, was measured using optical backscatter reflectometry to be 1.14 dB/m. To realize high Q resonators and calibrate the simulated coupling coefficients, we designed coupling test structures with a straight waveguide and a bent waveguide with a bend radius of 11.83mm. The gap distance varied from 1.5 μm to 5.5 μm. The simulated and measured coupler data were used to realize two different designs of laser resonators. The first resonator has a radius of 11.83 mm and is 74.3 mm long. The resonator coupler consists of a straight waveguide bus passing by the resonator with a spacing of 5.42 μm to achieve 0.5 % coupling. The bus waveguide also has a s-bend that offset the input and output from each other laterally by 200 μm. The bus waveguides taper to 5μm at the facet to reduce fiber to chip coupling losses. The second resonator design has all the same parameters except for the coupling gap and resonator radius which are 6 μm and 11.83 mm respectively to achieve 0.25 % coupling. This gives the second resonator a length of 74.1 mm.

**Pump probe Brillouin gain measurement:** We determined the Brillouin gain coefficient of the fabricated ultra-low loss $Si_3N_4$ coils using pump probe technique[31]. Two stable, tunable lasers were directed through circulators and polarization controllers, then coupled into a temperature stabilized 5 m long segment of waveguide from opposite ends. The silica fiber pigtail of the circulator that couples pump laser in to waveguide spiral was shortened and spliced to a segment of a highly nonlinear single mode Nufern UHNA3 fiber which has a Brillouin gain peak spectrally isolated from our waveguide. This minimized the contribution of probe gain within silica fiber whose gain spectrum is very close to $Si_3N_4$ waveguides used in this work. The pump was amplified with an erbium-doped fiber amplifier and the difference in optical frequency between the two lasers was swept through the anticipated position of the SBS gain peak. The beat note between the pump and probe was monitored by combining the two and monitoring the signal they generated in an electrical spectrum analyzer. As the pump and probe beams propagated counter to each other within the waveguide, the pump contributed gain to the probe based on the frequency-dependent value of the SBS gain coefficient. The probe gain was measured for on-chip pump powers varying from about 200 mW to 600 mW as the pump laser frequency was tuned to vary the spacing between the two lasers from about 10.7 GHz to 11.1 GHz. The maximum probe gain was plotted as a function of on-chip pump power and the peak SBS gain coefficient over the effective spiral length was extracted by performing a numerical fit of the data. The loss coefficient of the spiral, extracted using optical backscatter reflectometry to be 1.14 dB/m, was used to calculate the effective length of spiral as 2.7843 m. Using this data, the peak SBS gain coefficient is $0.10 \pm .009$ $m^{-1}W^{-1}$ which is close to the simulated value. The results gave us confidence in the accuracy of our theoretical methods for SBS gain calculation, and provided us inputs to design a Brillouin ring laser resonator.

**RF calibrated Mach-Zehnder interferometer based Q measurement:** A 200 m unbalanced fiber based Mach-Zehnder interferometer (MZI) was used as a reference frequency spectrum to accurately measure the resonator Q. A single side band swept laser source was used to calibrate the FSR of the MZI. The SSB frequency swept source was generated by passing an intensity modulated laser through cascaded stages of fiber Bragg gratings to suppress the carrier and upper sideband. The filtered lower sideband was passed through the MZI and the modulation frequency was swept in time using a microwave synthesizer. The interferometer FSR was measured by monitoring the transmitted power on an oscilloscope as the optical source frequency was swept. Transmitted intensity maxima (and minima) frequency spacing correspond to the interferometer FSR, measured to be $1.07 \pm .0146$ MHz by simultaneously scanning the laser frequency through both the MZI and device under test, the MZI fringe spacing (FSR) provides an RF calibrated frequency reference for accurate evaluation of resonator Q factors.

**Laser threshold measurement:** A low frequency noise tunable laser source was phase modulated at 40 MHz, amplified using an Erbium doped fiber amplifier and edge coupled in to our ring resonator. The laser frequency was tuned to the cavity mode of the temperature controlled ring resonator and locked using Pound-Drever-Hall technique. The optical transmission and reflection spectra of the locked resonator were recorded simultaneously using two high resolution optical spectrum analyzers for varying input pump powers. The on-chip optical power in each Stokes order was evaluated from the measured optical spectra. Even order Stokes powers were evaluated from the transmission spectra while odd orders powers were extracted from the reflection spectra of the resonator. A plot of on-chip Stokes order powers vs on-chip pump powers was used to evaluate the lasing threshold of each Stokes component.

**Data availability**

The data that support the plots within this paper and other findings of this study are available from the corresponding author on reasonable request.